\documentclass[aps,prl,showpacs, letterpaper,twocolumn,superscriptaddress]{revtex4}

\usepackage{amssymb}
\usepackage{amsmath}
\usepackage{latexsym}
\usepackage{graphicx}
\usepackage{placeins}
\usepackage{booktabs}
\usepackage{epstopdf}

\newcommand{\bvec}[1]{\mathbf{#1}}

\newcommand{\K}{\mathrm{K}}

\newcommand{\averageop}[3]{\left\langle#1\left\lvert#2\right\rvert#3\right\rangle}

\begin{document}

\title{Displaced path integral formulation for the momentum distribution of quantum particles}
\affiliation{Program in Applied and Computational Mathematics, Princeton
University, Princeton, NJ 08544}
\affiliation{Department of Chemistry, Princeton University, Princeton, NJ 08544}
\affiliation{Department of Chemistry, Princeton University, Princeton, NJ 08544}
\affiliation{Department of Physics, Princeton University, Princeton, NJ 08544}
\affiliation{Computational Science, Department of Chemistry and Applied
Biosciences, ETH Zurich, USI Campus, Via Giuseppe Buffi 12, CH-6900 Lugano, Switzerland}
\author{Lin Lin}
\affiliation{Program in Applied and Computational Mathematics, Princeton
University, Princeton, NJ 08544}
\author{Joseph A. Morrone}
\altaffiliation{Present address: Department of Chemistry, Columbia University, New York NY 10027}
\affiliation{Department of Chemistry, Princeton University, Princeton, NJ 08544}
\author{Roberto Car}
\email{rcar@princeton.edu}
\affiliation{Department of Chemistry, Princeton University, Princeton, NJ 08544}
\affiliation{Department of Physics, Princeton University, Princeton, NJ 08544}
\author{Michele Parrinello}
\affiliation{Computational Science, Department of Chemistry and Applied
Biosciences, ETH Zurich, USI Campus, Via Giuseppe Buffi 12, CH-6900 Lugano, Switzerland}


\begin{abstract}

The proton momentum distribution, accessible by deep inelastic
neutron scattering, is a very sensitive probe of the potential of
mean force experienced by the protons in hydrogen-bonded systems.
In this work we introduce a novel estimator for the end to end
distribution of the Feynman paths, i.e. the Fourier transform of
the momentum distribution. In this formulation, free particle and
environmental contributions factorize. Moreover, the
environmental contribution has a natural analogy to a free energy
surface in statistical mechanics, facilitating the interpretation
of experiments. The new formulation is not only conceptually but
also computationally advantageous.  We illustrate the method with
applications to an empirical water model, ab-initio ice, and one dimensional
model systems.

\end{abstract}

\maketitle

The behavior of protons and more generally of light nuclei  in condensed
phases is significantly affected by quantum effects even at ambient
temperatures. The isotopic effect in water, the ferroelectric behavior
of  KDP, and the formation of  high pressure ice phases, are just a few
of the relevant phenomena where the quantum behavior of the nuclei plays
a  role. To address these issues  a powerful experimental tool,  deep
inelastic neutron scattering (DINS) 
that
measures the  momentum distribution ~\cite{kdp,reiter,andreani}
has recently been developed.
Quantum effects are revealed by strong deviations from the classical
Maxwell distribution. However interpreting DINS experiments is difficult
and so far has been based on extensive  and challenging $\it{ab\ initio}$
molecular dynamics simulations~\cite{morrone08,morrone09}. While these
calculations  have shown that good agreement between theory and
experiments is possible, a simpler way of calculating the momentum
distribution needs to be found and  the link between the experimental
data and the underlying physics made  transparent  if DINS is to become
a standard tool.
 
In order to understand the source of this computational challenge, let us
contrast the expression  for the momentum distribution
$n(\bvec{p})$ and that of the partition function $Z$ in terms of the
density matrix $\rho(\bvec{r},\bvec{r}')=\averageop{\bvec{r}}{e^{-\beta
H}}{\bvec{r}'}$. The former may be expressed as: 
\begin{equation}
  \begin{split}
  n(\bvec{p}) &= \frac{1}{(2\pi\hbar)^3 Z}\int d\bvec{r}d\bvec{r}'
  e^{\frac{i}{\hbar}\bvec{p}\cdot(\bvec{r}-\bvec{r}')}
  \rho(\bvec{r},\bvec{r}') \\
  &= \frac{1}{(2\pi\hbar)^3}\int
  d\bvec{x} e^{\frac{i}{\hbar}\bvec{p}\cdot \bvec{x}}
  \widetilde{n}(\bvec{x})
  \end{split}
  \label{}
\end{equation}
where  $\widetilde{n}(\bvec{x})=\frac{1}{Z}\int
d\bvec{r} d\bvec{r}'\delta \left(\bvec{r}-\bvec{r}'-\bvec{x}\right) 
\rho\left(\bvec{r},\bvec{r}'\right)$.
The partition function is given by:
\begin{equation}
  Z=\int d\bvec{r}\; \rho \left(\bvec{r},\bvec{r}\right).
   \label{}
\end{equation}
It can be seen that $n(\bvec{p})$ involves the off-diagonal matrix
elements while $Z$ is determined solely by diagonal terms. In a
condensed system the potential energy surface in which the particles
move
is in a high dimensional space and statistical sampling is the only viable
computational strategy. This is usually done using the Feynman path
representation.  In this representation, $\widetilde{n}(\bvec{x})$    is
an end to end distribution of a sum over open paths,  while closed
ones determine  $Z$~\cite{ceperley,ceperley2}. Sampling  is done on the closed paths that
specify $Z$ and it is challenging from these simulations to estimate
the open path distribution that determines $n(\bvec{p})$.

One approach is to artificially open a fraction of the
paths~\cite{morrone07}.  In so doing one has to balance two contradictory
requirements. On one hand the number of open paths has to be large enough
to obtain good statistics for
$\widetilde{n}(\bvec{x})$, while on the other hand it cannot be too large
as the sampling will become incorrect. In this work
we introduce a new expression for $\widetilde{n}(\bvec{x})$
which does not require opening the paths and compromises neither
sampling accuracy nor statistics. Following a derivation whose detail
can be found in the supplementary material we find: 
\begin{widetext}
\begin{equation}
  \widetilde{n}(\bvec{x})=\widetilde{n}_0(\bvec{x})
  \dfrac{\int\mathfrak{D}\bvec{r}(\tau) 
  \exp\left(-\frac{1}{\hbar} 
  \int_{0}^{\beta\hbar}d\tau\; \left(\frac{m\dot{\bvec{r}}^2(\tau)}{2} + 
  V[\bvec{r}(\tau)+y(\tau)\bvec{x}] \right)\right)}{\int\mathfrak{D}\bvec{r}(\tau) 
  \exp\left(-\frac{1}{\hbar} 
  \int_{0}^{\beta\hbar}d\tau\; \left(\frac{m\dot{\bvec{r}}^2(\tau)}{2} + 
  V[\bvec{r}(\tau)]\right)\right)}, 
  \label{eqn:nxsingle}
\end{equation}
\end{widetext}
where $\widetilde{n}_{0}(\bvec{x})=e^{-\frac{m\bvec{x}^{2}}{2\beta \hbar
^{2} } } $ is the free particle end to end distribution. The function
$y(\tau)$ is arbitrary but for the boundary
condition $y(\beta\hbar)-y(0)=1$. In practice,
the optimal choice is to take $y=\frac{1}{2}
-\frac{\tau }{\beta\hbar }$ since it minimizes the distance between
$\bvec{r}(\tau)$ and the displaced path $\bvec{r}(\tau)+y(\tau)\bvec{x}$. Notice that, for simplicity, 
Eq.~\eqref{eqn:nxsingle} refers to a single particle subject to the external potential $V[\bvec{r}]$.   
Generalization to many-body systems is straightforward if exchange effects between identical particles can be neglected.  How to include  
such effects will be discussed in a future publication. Eq.~\eqref{eqn:nxsingle} merits
further comment. In the calculation of the kinetic energy it has been found
to be extremely useful to use estimators in which the free particle
contribution has been
explicitly accounted for~\cite{virial}. We expect similar computational
advantages from the explicit separation of
$\widetilde{n}_{0}(\bvec{x})$. Furthermore it follows from
Eq.~\eqref{eqn:nxsingle} that, having put $Z(\bvec{0})=Z$, we can
write $\frac{\widetilde{n}(\bvec{x})}{\widetilde{n}_0(\bvec{x})} 
=\frac{Z(\bvec{x})}{Z(\bvec{0})} $
as a ratio between  two partition functions.
To calculate this ratio or its logarithm $U(\bvec{x})=-\ln
\frac{Z(\bvec{x})}{Z(\bvec{0})} $ standard statistical mechanics methods
such as  free energy perturbation~\cite{Zwanzig1954} or thermodynamic
integration~\cite{kirkwood1935} may be utilized.

Using  free energy perturbation one may compute:
\begin{equation}
  U(\bvec{x})=-\ln \left\langle e^{
  -\frac{1}{\hbar}\int_{0}^{\beta\hbar}d\tau\;
  \left(V[\bvec{r}(\tau)+y(\tau)\bvec{x}]-V[\bvec{r}(\tau)]\right) }
  \right\rangle_{\bvec{0}}.
  \label{eqn:closeformula}
\end{equation}
where the average is evaluated using  the closed path distribution $Z(\bvec{0})$.

The free energy perturbation method can only be applied to
systems with weak quantum effects. For systems with strong quantum
effects the average is difficult to converge and instead we use
thermodynamic integration.  In this scheme  $U(\bvec{x})$ is obtained as
an
integral $U(\bvec{x})=\int_{0}^{\bvec{x}}d\bvec{x}'\cdot
\bvec{F}(\bvec{x}')$ over the mean force,
\begin{equation}
  \bvec{F}(\bvec{x}') = \left\langle
  \frac{1}{\hbar}\int_{0}^{\beta\hbar} d\tau\; \nabla_{\bvec{r}}V[\bvec{r}(\tau)+
  y(\tau)\bvec{x}']y(\tau)\right\rangle_{\bvec{x}'}
  \label{eqn:TI}
\end{equation}
evaluated at the intermediate distributions $Z(\bvec{x}')$.
In this case  thermodynamic integration
 requires opening the paths, but it does so in  a fully controlled
way. 
Besides being rigorous our estimator  offers several computational
advantages.  In three dimensions the standard approach suffers from poor
statistics at short distances  due to the geometrical $r^2$
factor, and this is not the case here. By averaging over all the
particles, statistics can be greatly improved. The calculation over different
particles is intrinsically parallel and the power of modern computers
optimally harnessed.  Furthermore, in crystals where anisotropies are
relevant, the dependence of $n(\bvec{p})$ on the momentum direction can be
easily evaluated. 

\begin{figure}[h]
    \begin{center}
      \includegraphics[width=3in]{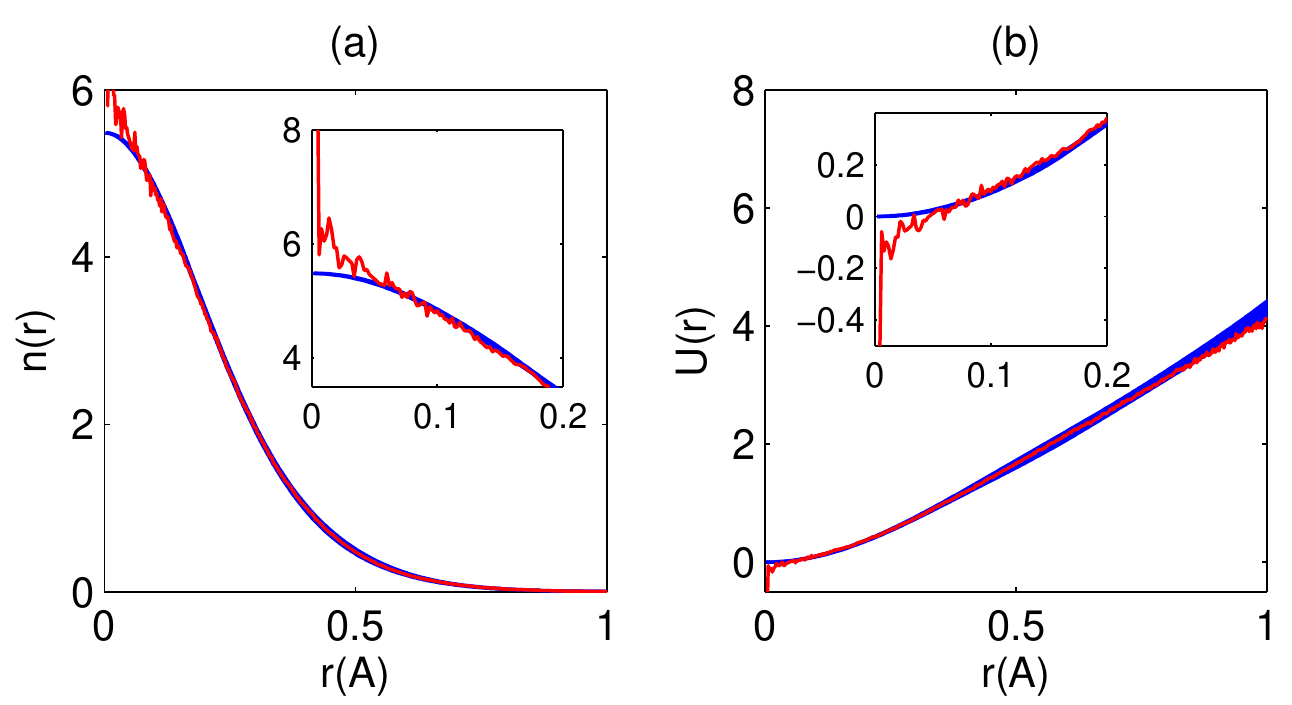}
    \end{center}
    \caption{(color online) Comparison of (a) the end to end
    distribution and (b) the potential of mean force in SPC/F2 water.
    In both figures, the red line is computed by a
    $268$ps open path integral simulation.
    The thick blue line is calculated using the displaced path estimator
    \eqref{eqn:closeformula}, with the thickness indicating the
    $95\%$ confidence interval. The noise near $r=0$ in both insets for
    open path simulation is due to the $r^2$ weight in the spherical
    integration, while
    the displaced path gives correct small $r$ behavior by definition.} 
    \label{fig:excesswater} 
  \end{figure}

We first test our algorithm on a
flexible model for water~\cite{LobaughVoth1997}.
The simulation box contains $32$ water molecules. The temperature is set to
be $296\K$. Both protons and oxygens are treated by quantum mechanics,
and are represented by $64$ classical beads. The end to end distribution
is spherically averaged in water. The quantum effect for water at room
temperature is relatively small~\cite{morrone08}. This
allows us to use free energy perturbation  
\eqref{eqn:closeformula} and compare the results with open
path integral simulation~\cite{morrone07}. In
the latter case, in principle one proton path should be opened and all
other paths should be closed.  However, the resulting statistics would
be poor. In order to boost statistics one proton path per water molecule
was opened, as it was found that this approximation leads to a
negligible error in the momentum distribution due to the relatively weak
interaction between protons belonging to different water
molecules~\cite{morrone07}.  The closed path formulation  allows one to
compute the end to end distribution without
opening any proton path, and therefore all the protons can be included in 
the calculation of the end to end distribution without any approximation.  
We show the end to end distribution
calculated both from a $268$~ps open path simulation and from a $12$~ps closed path
simulation that utilizes the estimator given by
Eq.~\eqref{eqn:closeformula} in Fig.~\ref{fig:excesswater} (a), and the
comparison of the potential of mean force in Fig.~\ref{fig:excesswater} (b). In both simulations, the
time step is $0.24$~fs.  Two consecutive steps contain
highly correlated information, and the free energy perturbation estimator 
may be computed every $20$ steps. Thus with only a small increase in
computational overhead in comparison 
to an open path simulation of the same length, the displaced path
formulation has a large gain in terms of sampling this property
efficiently.

The thermodynamic integration approach given in Eq.~\eqref{eqn:TI} is
not only computationally advantageous but also provides one with the
potential of
mean force $U(\bvec{x})$, and its gradient  $\bvec{F}(\bvec{x})$ which are key quantities
for interpreting the physics underlying $n(\bvec{p})$.  We first note
that the kinetic energy $K$ is given by $K = \frac{\hbar^2}{2m}
\nabla \cdot \bvec{F}(\bvec{x})\Big\vert_{\bvec{x}=\bvec{0}} +
\frac{3}{2\beta} \equiv K_V + \frac{3}{2\beta}$. Since $3/2\beta$ is
the free particle contribution, the non-classical contribution is
completely included in the excess kinetic energy term $K_V$, and is
determined by the zero point curvature of $U(\bvec{x})$.  Secondly, if the momentum
distribution of an individual particle is accessible (as is possible e.g. in simulations) and the underlying
potential energy surface is harmonic, the
end to end distribution follows a Gaussian distribution and the
mean force is given by a straight line.  Any deviation of
$\hat{\bvec{q}} \cdot \bvec{F}(\bvec{x})$ from linearity signals anharmonic behavior along
the $\hat{\bvec{q}}$ direction. 

\begin{figure}[h]
  \begin{center}
    \includegraphics[width=2in]{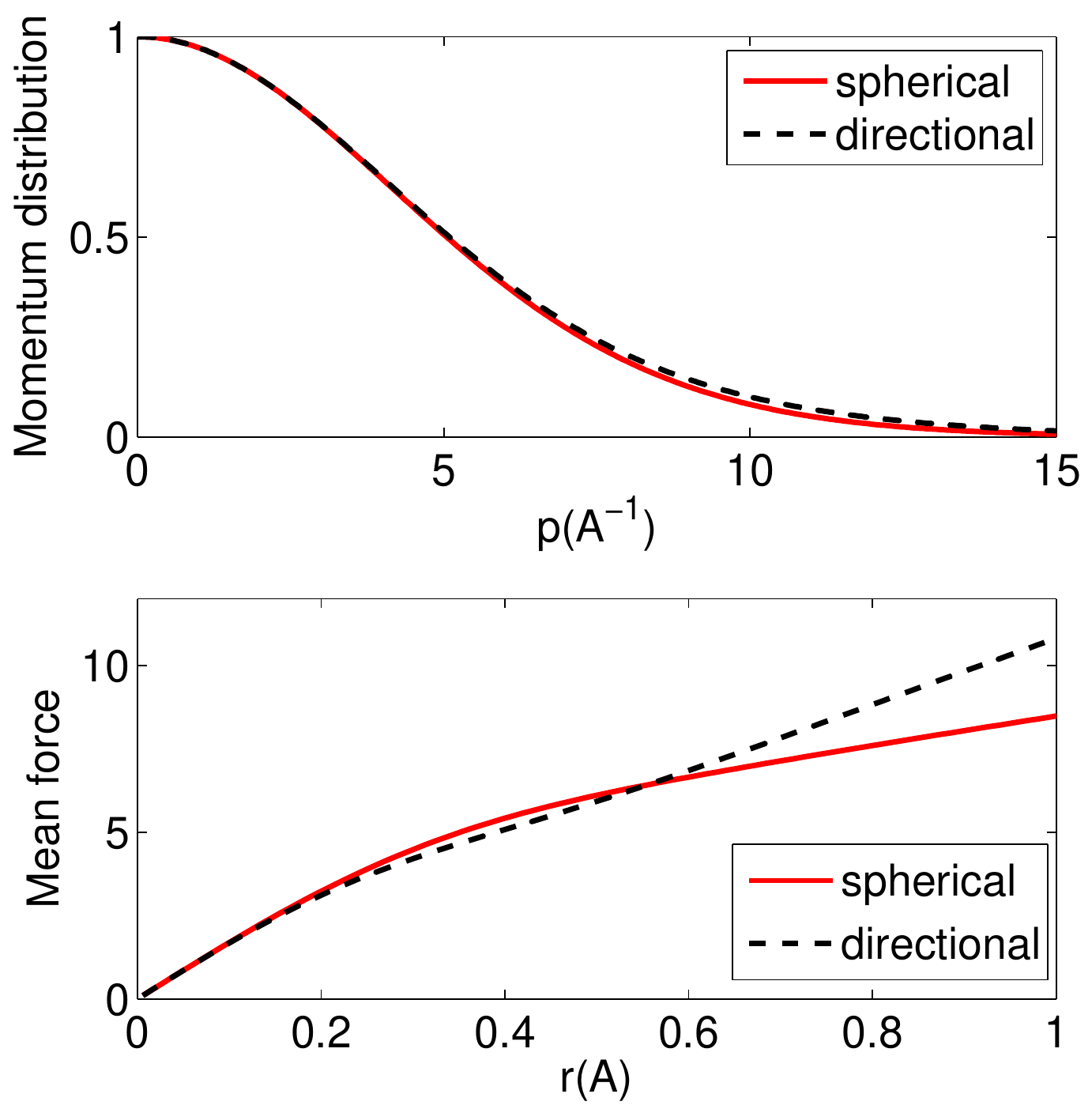}
  \end{center}
  \caption{(color online) Top panel: the momentum distribution of the
  protons in ice Ih resulting from an anisotropic harmonic model (see
  text). Both the spherical and the directional distribution along the
  c-axis are shown. Bottom panel: the corresponding spherical and
  directional mean force projected along the c-axis.  The curves are
  plotted as a function of the end to end distance.  The mean force enhances the
  differences between spherical and directional distributions.}
  \label{fig:npforce} 
\end{figure}

In experiments, the spherically averaged momentum distribution is accessible 
in liquids, and amorphous and polycrystalline solids, while
the directional distribution is also accessible in mono
crystalline materials. The latter distribution provides more information about the underlying
potential energy surface. However, in single crystals the total
momentum distribution is the sum of the contributions of 
individual particles participating in bonds with different
orientations. As a consequence the difference between directional
and spherical momentum distribution is usually very small as shown
in the top panel of Fig.~\ref{fig:npforce}. This figure is based on an
anisotropic harmonic model~\cite{icestat} with three distinct principal
frequencies that is fit to the ab initio path
integral data for ice Ih~\cite{morrone08}.  The bottom panel of the same figure
clearly shows that the distinction between the spherical and directional distributions is enhanced when comparing
the mean forces.  It is therefore of great interest
to link directly the mean force to the experimental data, i.e. to the
Compton profile $J(\hat{\bvec{q}},y)=\int n(\bvec{p}) \delta
(y-\bvec{p}\cdot \hat{\bvec{q}}) d\bvec{p}$ where $\hat{\bvec{q}}$ indicates
the direction of the neutron detector~\cite{reiter}. One finds with a
derivation provided in the supplemental material that the mean force is
related to the Compton profile by:
\begin{equation}
  \hat{\bvec{q}}\cdot \bvec{F}(x\hat{\bvec{q}}) =
  -\frac{mx}{\beta\hbar^2}+\frac{\int_{0}^{\infty}dy\; y
  \sin(xy/\hbar)J(\hat{\bvec{q}},y)}{\hbar\int_{0}^{\infty}dy\; 
  \cos(xy/\hbar)J(\hat{\bvec{q}},y)}.
  \label{eqn:FJ}
\end{equation}
In the bottom panel of Fig.~\ref{fig:npforce} the slope of the mean
force, either spherical or directional, at $r=0$ is equal to the excess
kinetic energy $K_V$ divided by the constant $\frac{\hbar^2}{2m}$. This 
is an exact result that originates from the symmetry property of ice Ih. In
general the spherical and directional mean force can have different
slopes at $r=0$. The
deviation of the spherical and directional forces from linearity at finite
$r$ results from the averaging process and is not a sign of
anharmonicity.  Thus in the interpretation of the experimental Compton
profile, which results from the contribution of many particles, one must
distinguish the case of an anisotropic harmonic potential energy surface
from that of an anharmonic potential energy surface.  To the best of our
knowledge the procedure that is currently adopted to fit the
experimental data~\cite{reiter,andreani,supercool} does not separate
well anisotropic and anharmonic effects.  We propose here an alternative
approach in which the mean force is associated to the experimental
Compton profile according to Eq.~\eqref{eqn:FJ}.  The projections of the
mean force along different directions are then fitted to an anisotropic
harmonic model averaged as required by the crystal symmetry.  Any systematic
deviation from experiment of the mean force originating from the
harmonic contribution, can then be associated  to
anharmonicity and used to further refine the underlying model potential
energy surface.

\begin{figure}[h]
  \begin{center}
    \includegraphics[width=3in]{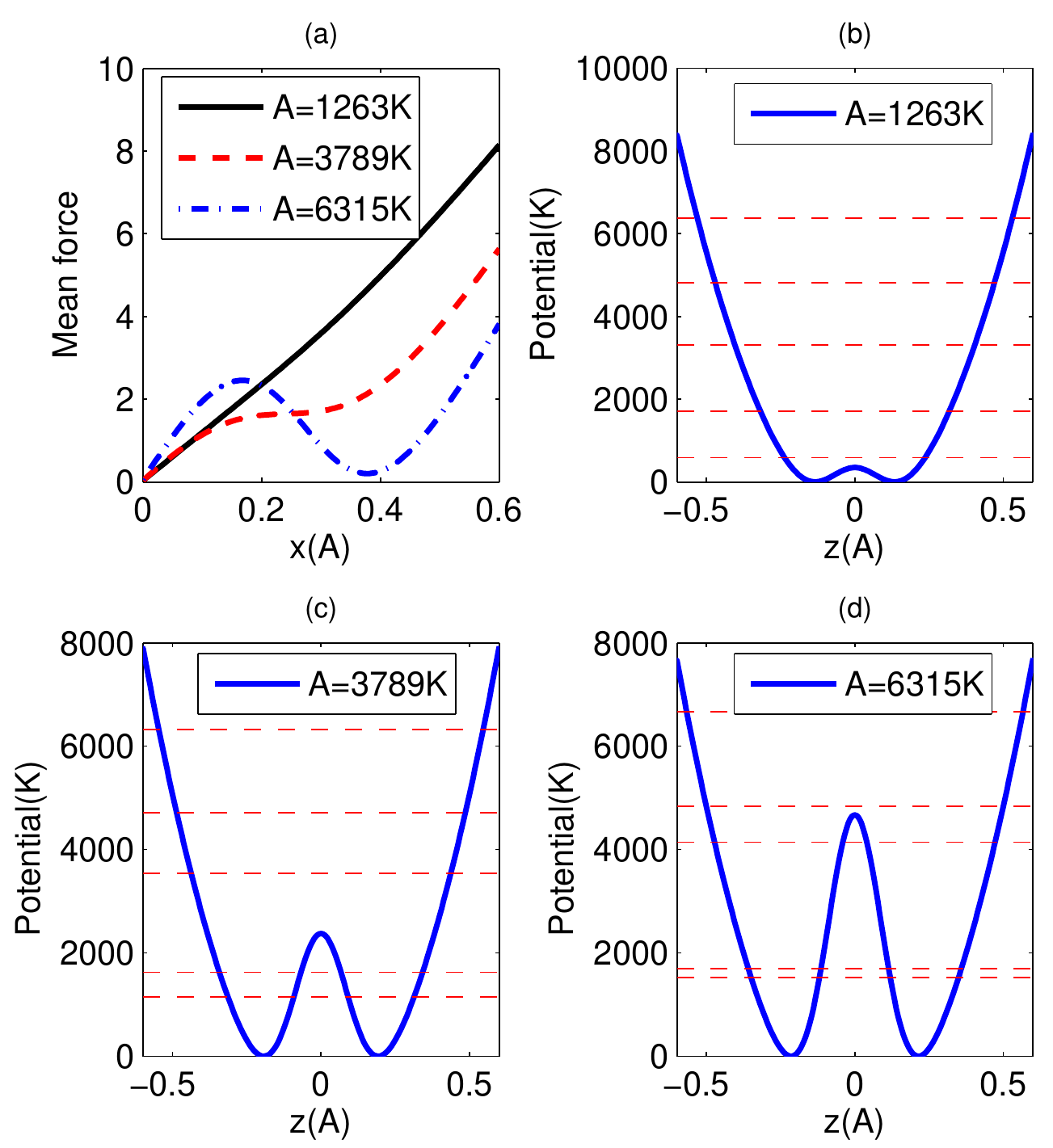}
  \end{center}
  \caption{(color online) (a) The mean force corresponding to a
  double well model at $T=30\K$, for different barrier heights
  $A=1263\K$ (black solid line), $A=3789\K$ (red dashed line), and
  $A=6315\K$ (blue dot-dashed line). (b) Potential energy surface for
  $A=1263\K$ (blue solid line), and the first five energy levels (red
  dashed line). (c) (d) the same as (b), but with $A=3789\K$ and
  $A=6315\K$ respectively.}
  \label{fig:gf3A} 
\end{figure}

The framework introduced here may be also utilized to provide insight to
the investigation of anharmonic systems.
Consider for example a particle with the proton mass subject to a
model double well 1D-potential.
$V=\frac{m\omega^2}{2}z^2+A\exp(-\frac{z^2}{2\xi^2})$ with
$\omega=1578\K$, and
$\xi=0.094$\AA. $A$ characterizes the barrier height and is set to
be $1263\K, 3789\K$, and $6315\K$, respectively. These parameters mimic
different tunneling regimes for protons along a hydrogen bond~\cite{icecpmd,morrone09}. The temperature is set to be $30\K$.  At this
temperature the behavior of the systems is dominated by the ground-state,
and the end to end distribution can be approximated by the overlap
integral $\widetilde{n}(x)=\int d z \psi(z) \psi(z+x)$ where $\psi(z)$
is the ground-state wavefunction and $F(x) = -\frac{d}{dx}\ln
\widetilde{n}(x)$.  In Fig.~\ref{fig:gf3A}   we can see how
qualitatively different the mean force can be in the three cases. One goes
from a fully monotonic behavior for $A=1263\K$ which is a model for a low
energy barrier hydrogen bond~\cite{benoit05}, to the strongly non
monotonic mean forces for $A=3789\K, A=6315\K$ where the tunneling 
states lie below the barrier height. Additionally, it is not very difficult
to relate features of the mean force to the underlying effective
potential. 

\begin{figure}[h]
  \begin{center}
    \includegraphics[width=2in]{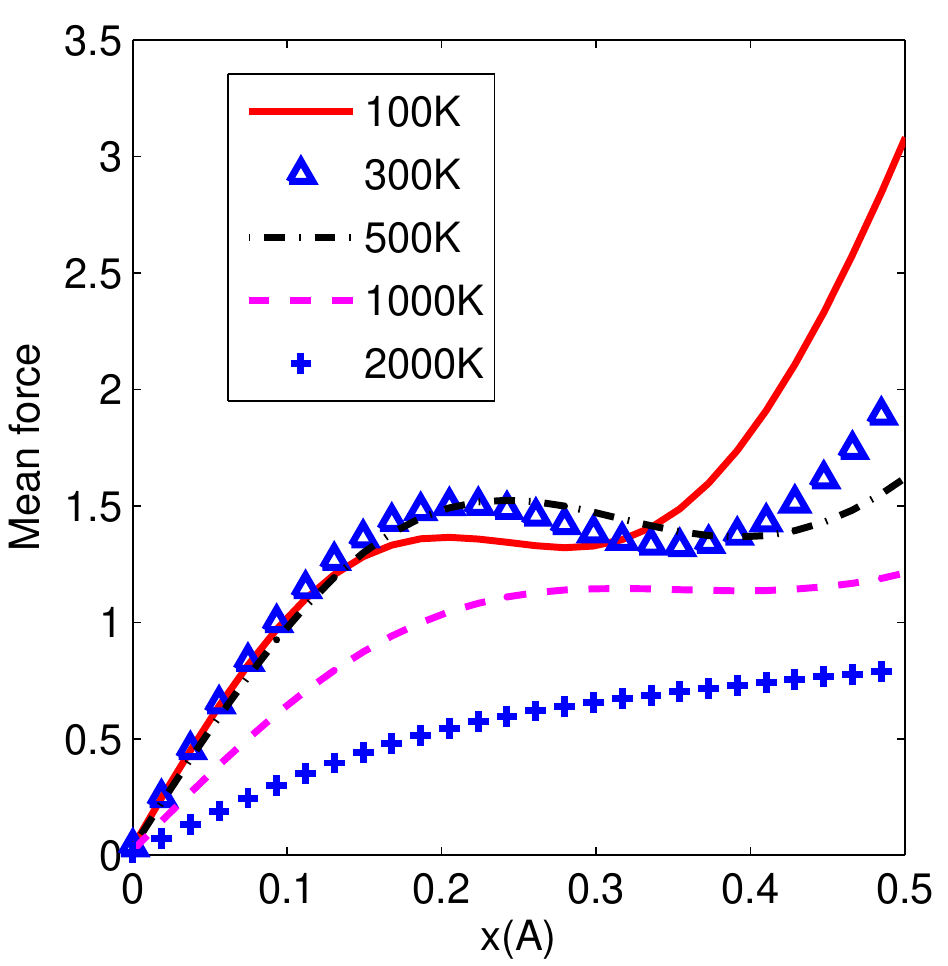}
  \end{center}
  \caption{(color online) The mean force corresponding to a
  double well model at $A=3789\K$ for different temperatures $100\K$
  (red solid line), $300\K$ (blue triangle), $500\K$ (black dot-dashed line),
  $1000\K$ (magenta dashed line), and $2000\K$ (blue cross). }
  \label{fig:gf5T} 
\end{figure}

It is also instructive to study $F(x)$   as a function of temperature
when the higher states are mixed in the density matrix.  This is done in
Fig.~\ref{fig:gf5T}  for the double well potential with  $A=3789\K$.  
For temperatures in the $100-500\K$ range, the behavior is dominated by the
two lowest eigenstates. The slope of $F(x)$ at small $x$, which is
proportional to the excess kinetic energy $K_V$, shows little dependence on
$T$.  It can be shown with detailed analysis that this is a generic
feature of two level tunneling systems. Other characters seen in
Fig.~\ref{fig:gf5T} in the same range of temperatures, such as the more
pronounced kink at intermediate $x$ and the enhanced softening of the mean
force at large $x$, derive from the odd symmetry of the first excited
state contribution.  Eventually at higher $T$ the kink in $F(x)$ disappears
as the mean force progressively resumes linear behavior with a
slope that tends to zero as high temperature classical limit is reached.


In this work, we develop a novel displaced path formalism for the
calculation of momentum distribution of quantum particles.  The
algorithm is rigorous  and computationally advantageous.  The new
formulation introduces in a natural way a potential of mean force which
is a quantity that very clearly illuminates the physics behind
$n(\bvec{p})$ and can be used to further understand and analyze
experimental and theoretical results.

This work is partially supported by NSF under grant CHE-0956500 and by DOE
under grant DE-FG02-05ER46201 (LL and RC).

\end{document}


\title{Supplemental material for displaced path integral formulation for the momentum distribution of quantum particles}
\affiliation{Program in Applied and Computational Mathematics, Princeton
University, Princeton, NJ 08544}
\affiliation{Department of Chemistry, Princeton University, Princeton, NJ 08544}
\affiliation{Department of Chemistry, Princeton University, Princeton, NJ 08544}
\affiliation{Department of Physics, Princeton University, Princeton, NJ 08544}
\affiliation{Computational Science, Department of Chemistry and Applied
Biosciences, ETH Zurich, USI Campus, Via Giuseppe Buffi 12, CH-6900 Lugano, Switzerland}
\author{Lin Lin}
\affiliation{Program in Applied and Computational Mathematics, Princeton
University, Princeton, NJ 08544}
\author{Joseph A. Morrone}
\altaffiliation{Present address: Department of Chemistry, Columbia University, New York NY 10027}
\affiliation{Department of Chemistry, Princeton University, Princeton, NJ 08544}
\author{Roberto Car}
\email{rcar@princeton.edu}
\affiliation{Department of Chemistry, Princeton University, Princeton, NJ 08544}
\affiliation{Department of Physics, Princeton University, Princeton, NJ 08544}
\author{Michele Parrinello}
\affiliation{Computational Science, Department of Chemistry and Applied
Biosciences, ETH Zurich, USI Campus, Via Giuseppe Buffi 12, CH-6900 Lugano, Switzerland}

\maketitle

Derivation of Eq.~(3) in the text:

Within Feynman's path integral representation the density operator is
given by:
\begin{equation}
  \rho(\bvec{r},\bvec{r}') = \int_{\bvec{r}(0)=\bvec{r},
\bvec{r}(\beta\hbar)=\bvec{r}'} \mathfrak{D} \bvec{r}(\tau)
e^{-\frac{1}{\hbar}\int_{0}^{\beta\hbar} d\tau\;\left(
\frac{m\dot{\bvec{r}}^2(\tau)}{2}+V[\bvec{r}(\tau)]\right)},
  \label{}
\end{equation}
and the end-to-end distribution is:
\begin{equation}
  \begin{split}
  \widetilde{n}(\bvec{x}) = &\frac{1}{Z}\int d\bvec{r} d\bvec{r}'
  \delta\left(\bvec{r}-\bvec{r}'-\bvec{x}\right)
  \rho\left(\bvec{r},\bvec{r}'\right)\\
  = & \frac{\int_{\bvec{r}(0)-\bvec{r}(\beta\hbar)=\bvec{x}}
  \mathfrak{D} \bvec{r}(\tau) e^{-\frac{1}{\hbar}\int_{0}^{\beta\hbar}
  d\tau\;
  \left(\frac{m\dot{\bvec{r}}^2(\tau)}{2}+V[\bvec{r}(\tau)]\right)}}{ 
  \int_{\bvec{r}(\beta\hbar)=\bvec{r}(0)}
  \mathfrak{D} \bvec{r}(\tau) e^{-\frac{1}{\hbar}\int_{0}^{\beta\hbar}
  d\tau\;
  \left(\frac{m\dot{\bvec{r}}^2(\tau)}{2}+V[\bvec{r}(\tau)]\right)}}.
  \end{split}
  \label{eqn:open}
\end{equation}

We now perform a linear transformation in path space in the expression
on the numerator:
\begin{equation}
  \bvec{r}(\tau) = \widetilde{\bvec{r}}(\tau) + y(\tau)\bvec{x}.
  \label{eqn:dp}
\end{equation}
Here $y(\tau)=C-\frac{\tau}{\beta\hbar}$ and $C$ is an arbitrary real
number.  Then the numerator is given by
\begin{equation}
  \begin{split}
    &\int_{\bvec{r}(0)-\bvec{r}(\beta\hbar)=\bvec{x}}
  \mathfrak{D} \bvec{r}(\tau) e^{-\frac{1}{\hbar}\int_{0}^{\beta\hbar}
  d\tau\;
  \left(\frac{m\dot{\bvec{r}}^2(\tau)}{2}+V[\bvec{r}(\tau)]\right)}\\
  =& e^{-\frac{m\bvec{x}^{2}}{2\beta \hbar ^{2} } } 
  \int_{\widetilde{\bvec{r}}(\beta\hbar)=\widetilde{\bvec{r}}(0)}
  \mathfrak{D} \widetilde{\bvec{r}}(\tau) e^{-\frac{1}{\hbar}\int_{0}^{\beta\hbar}
  d\tau\;\left(
  \frac{m\dot{\widetilde{\bvec{r}}}^2(\tau)}{2}+V[\widetilde{\bvec{r}}(\tau)+y(\tau)\bvec{x}]\right)}.
  \end{split}
  \label{}
\end{equation}
Eq.~\eqref{eqn:dp} transforms the open path $\bvec{r}(\tau)$ into the
closed path $\widetilde{\bvec{r}}(\tau)$, and the free
particle contribution comes naturally from the derivative of $y(\tau)$.
The choice of the constant $C$ influences the variance of free
energy perturbation and thermodynamic integration
estimators in the text.
It is found that the lowest variance is achieved when $C=1/2$,
since this choice has the smallest displacement from the closed path
configuration.   This is Eq.~(3) in the text.

Next we present the derivation of Eq.~(6) in the text:

The Compton profile is given by
\begin{equation}
 J(\hat{\bvec{q}},y)=\int n(\bvec{p}) \delta (y-\bvec{p}\cdot
 \hat{\bvec{q}}) d\bvec{p}.
\end{equation}
The direction $\hat{\bvec{q}}$ is defined by the experimental setup, 
and the momentum distribution $n(\bvec{p})$ can be expressed in terms of
the end-to-end distribution $\widetilde{n}(\bvec{x})$ as
\begin{equation}
  n(\bvec{p})=\frac{1}{(2\pi\hbar)^3}\int
  d\bvec{x} e^{\frac{i}{\hbar}\bvec{p}\cdot \bvec{x}}
  \widetilde{n}(\bvec{x}).
  \label{}
\end{equation}

We indicate by $x_{\parallel}=\bvec{x}\cdot \hat{\bvec{q}}$, and
$\bvec{x}_{\bot}$ the $\bvec{x}$ component orthogonal to $\hat{\bvec{q}}$.
Correspondingly $p_{\parallel}=\bvec{p}\cdot \hat{\bvec{q}}$, and
$\bvec{p}_{\bot}$ is the $\bvec{p}$ component orthogonal to $\hat{\bvec{q}}$.
One has
\begin{equation}
  \begin{split}
    J(\hat{\bvec{q}},y) & = \frac{1}{(2\pi\hbar)^3} \int
    d\bvec{x} d\bvec{p}\; \widetilde{n}(\bvec{x})
    e^{\frac{i}{\hbar}\bvec{p}\cdot \bvec{x}} \delta(y-\bvec{p}\cdot
    \hat{\bvec{q}}) \\
    & = \frac{1}{(2\pi\hbar)^3} \int
    dx_{\parallel} d\bvec{x}_{\bot} dp_{\parallel} d\bvec{p}_{\bot}\; \widetilde{n}(\bvec{x})
    e^{\frac{i}{\hbar}x_{\parallel}p_{\parallel} +
    \frac{i}{\hbar}\bvec{p}_{\bot}\cdot \bvec{x}_{\bot}}
    \delta(y-p_{\parallel}) \\
    &= \frac{1}{2\pi\hbar} \int
    dx_{\parallel}\; \widetilde{n}(x_{\parallel} \hat{\bvec{q}})
    e^{\frac{i}{\hbar}x_{\parallel}y}.
  \end{split}
\end{equation}
Given the end to end distribution can be expressed as
\begin{equation}
  \widetilde{n}(\bvec{x}) = e^{-\frac{m\bvec{x}^2}{2\beta\hbar^2}}
  e^{-U(\bvec{x})},
\end{equation}
the potential of mean force $U(\bvec{x})$ can be obtained from the
Compton profile as
\begin{equation}
  U(x_{\parallel}\hat{\bvec{q}}) = -\frac{m
  x_{\parallel}^2}{2\beta\hbar^2} - \ln \int dy\; J(\hat{\bvec{q}},y)
  e^{-\frac{i}{\hbar}x_{\parallel} y}.
\end{equation}
The mean force $\bvec{F}(\bvec{x})$ is the gradient of
$U(\bvec{x})$. Taking into account that $J(\hat{\bvec{q}},y)$ is an even
function of $y$ one obtains
\begin{equation}
  \hat{\bvec{q}}\cdot \bvec{F}(x_{\parallel}\hat{\bvec{q}}) =
  -\frac{mx_{\parallel}}{\beta\hbar^2}+\frac{\int_{0}^{\infty}dy\; y
  \sin(x_{\parallel}y/\hbar)J(\hat{\bvec{q}},y)}{\hbar\int_{0}^{\infty}dy\; 
  \cos(x_{\parallel}y/\hbar)J(\hat{\bvec{q}},y)}.
\end{equation}
This is Eq.~(6) in the text.
